# Countermeasures against Bernstein's Remote Cache Timing Attack

Janaka Alawatugoda, Darshana Jayasinghe, and Roshan Ragel, *Member, IEEE*

*Abstract*- Cache timing attack is a type of side channel attack where the leaking timing information due to the cache behaviour of a crypto system is used by an attacker to break the system. Advanced Encryption Standard (AES) was considered a secure encryption standard until 2005 when Daniel Bernstein claimed that the software implementation of AES is vulnerable to cache timing attack. Bernstein demonstrated a remote cache timing attack on a software implementation of AES. The original AES implementation can methodically be altered to prevent the cache timing attack by hiding the natural cache-timing pattern during the encryption while preserving its semantics. The alternations while preventing the attack should not make the implementation very slow. In this paper, we report outcomes of our experiments on designing and implementing a number of possible countermeasures.

## I. INTRODUCTION

Information security plays a major role in every computerized system today. Information of a particular institution should be stored and transferred securely without allowing unauthorized parties to access or modify them. At this instance, the concept of 'encryption' comes into the arena to keep secret information in unreadable manner even if the intruders acquire the system or the message. Encryption can be informally defined as the original data (known as plain text) is converted into an unreadable or meaningless data (known as cipher text) using an algorithm and a secret key. Therefore, even if an intruder steals the secret information (the converted data), they are unreadable without the secret key (generally, the algorithm is of public knowledge).

Advanced Encryption Standard (AES) is such an encryption algorithm. It has become the US Federal Standard for information security after DES, Data Encryption Standard, which has become breakable. AES is an evolved version of the Rijndael algorithm developed by John Daemen and Vincent Rijmen [1]. AES uses a fixed block size of 128-bit (16 bytes) and a key of size 128-bit, 192-bit or 256-bit.

Janaka Alawatugoda is with the Department of Statistics and Computer Science, Faculty of Science, University of Peradeniya, Peradeniya 20400, Sri Lanka (email : araliyauops06207@gmail.com)
Darshana Jayasinghe is with the Department of Computer Engineering, Faculty of Engineering , University of Peradeniya, Peradeniya 20400, Sri Lanka (email : darshana@ce.pdn.ac.lk)
Roshan Ragel is with the Department of Computer Engineering, Faculty of Engineering , University of Peradeniya, Peradeniya 20400, Sri Lanka (email : ragelrg@gmail.com)

According to the size of the key, the number of rounds for encryption is varied. They are 10, 12 or 14 rounds for 128-bit, 192-bit or 256-bit key respectively. In each round except the final round, four operations are taken part. They are *Sub Bytes*, *Shift Rows*, *Mix Columns* and *Add Round Key*. Byte arrays of size 4x4 (16 bytes) are used for each of these operations. After a particular number of rounds according to the size of key, the plain text is converted into the cipher text.

The main feature of side channel attacks is that they do not focus on breaking the cryptographic algorithm through algorithmic weaknesses of it [2]. Instead, they use leaking information from the cryptographic system. There are various kinds of side channels available: leaking timing information, electromagnetic radiation, acoustic signals, visual or light signals, and power consumption are the major ones. By carefully gathering and analysing those leaking information an attacker will be able to extract the secret information from a system. Since this is not based on mathematically breaking the cryptographic algorithm and it is very difficult to fully stop the information leakage of the crypto system, side channel attacks have become a huge threat for the security of the information.

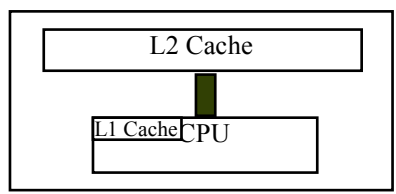

Fig. 1. Location of the L1 cache on a Processor Chip (CPU)

Cache Timing Attacks are a kind of side channel attacks, which uses leaking cache timing information as the side channel. At the time of execution, variables, data structures and other memory elements used for a particular program are loaded into the main memory (RAM). Cache memory is a high-speed memory, which is located in the processor for making the memory access faster through spatial and temporal locality as shown in Fig 1 (L1 Cache). Typically, the recently accessed memory areas are loaded into the cache. Since its high cost, cache memory is limited in size and only a limited amount of data can be stored. When a program needs to read a memory word, cache hardware checks to see if the line needed is in the cache. If so a cache hit occurs, the request is satisfied from the cache and no memory request is sent to the main memory. A cache hit normally takes two clock cycles [3]. When the memory word that the processor is looking for is not found in the cache, a cache miss happens, where the data have to be taken from the main memory (or from L2 cache), and therefore it takes longer time than cache hits. The time difference due to cache hits and misses are used as leaking timing information from the crypto system to perform cache timing attack.

If an attacker can collect the information about cache hits and misses of a software implementation of AES and then analyse them, he can get an idea about the process inside the system. This process and the knowledge about the cryptographic algorithm can be used to deduce the secret key of a crypto system. If the secret key could be found then the secrecy of the information is lost. Therefore, the secret information is open to intruders and AES becomes useless.

In this paper, we use a cache timing attack proposed by Daniel Bernstein [4] as the base of our experiments. We have implemented and tested a number of countermeasures against the attack proposed by Bernstein and proved that the countermeasures are working against remote cache timing attack. We have also evaluated the overheads of the proposed countermeasures.

The rest of the paper is organised as follows: related work is presented in Section II. In Section III, we discuss an investigation on Bernstein's attack. Section IV has details on countermeasures we implemented against Bernstein's attack and Section V on performance impact of experimented countermeasures. In Section VI, we conclude the paper.

## II. RELATED WORK

In year 2005, Daniel Bernstein pointed out [4] that AES is still vulnerable for timing attacks. He used 850MHz Pentium III desktop computer running FreeBSD 4.8 as a network server. He used OpenSSL 0.9.7a AES implementation in his server. The complete AES secret key was extracted using a client machine, which is connected to the network. After successfully completing the attack, he stated that the same technique could be used to extract complete AES key from most complicated servers, which handle Internet data, although the attackers would need additional timing data to average out the effects of network delays. He also stated that this attack was not limited to Pentium III processors. As he has tested an AMD Athlon, an Intel Pentium III, an Intel Pentium M, an IBM PowerPC RS64 IV, and a Sun UltreSPARC III processors also show comparable level of OpenSSL timing vulnerability.

Kocher et al. [5] has stated that by carefully measuring the amount of time required to perform private key operations, attackers might be able to find fixed Diffe-Hellman exponents, factor RSA keys, and break other cryptosystems. He also stated that these kinds of attacks are computationally inexpensive and often required a known cipher text. He pointed out that actual systems are potentially at risk whenever attackers could get reasonably accurate timing measurements.

David Brumley and Dan Bosh [6] have stated that timing attacks could not only be used for weak computing devices like smart cards but also applicable to attack general software systems. They have done an experiment to extract the RSA private key from OpenSSL based web server in a local area network. Their result showed that even if two virtual machines (network server virtual machine for making decryption queries and secure virtual machine for storing RSA private key) were running in the same machine for more protection of RSA private key, the network server virtual machine can extract the RSA private key from the secure virtual machine.

Felten et al. [7] have pointed out that since browsers perform various forms of caching, the time required for operations depends on the user's browsing history and this time variations convey enough information to compromise user's privacy. They have claimed that these attacks could not be prevented from simple countermeasures. Therefore, they have described a way of re-engineering browsers. In their paper, they have mentioned one fascinating example. Say, there are three people called A, B and C. Suppose that A is surfing the web and visits both B's and C's websites. If A want to know whether B has visited C's web site recently what A has to do is as follows. First, A access C's website and picks a file that can be seen by anyone who accesses C's website. Now A has to determine whether this file is in B's web cache. Therefore, A has to write a Java applet that implements the attack and embeds it in his home page. When B visits A's web site that applet automatically downloads and runs in B's browser. The applet measures the time required to access that particular file in C's web site. If that time is less than some threshold, A can conclude that B has visited C's website recently. Turning off caching, alternating hits or miss performance and turning off Java and Java Scripts are some of their countermeasures.

Tromer et al. [8] have described several software based side channel attacks based on inter-process leakage through the state of the CPU's memory cache. This leakage reveals memory access patterns. The attacks allow unprivileged process to attack other process running in parallel on the same processor even if partitioning method are available. They have also demonstrated an extremely strong type of attack, which requires knowledge of neither the specific plain texts nor cipher text but works by merely monitoring the effect of the cryptographic process on the cache. Avoiding memory access, alternative lookup tables, data independent memory access pattern and application specific algorithmic masking are some of the countermeasures proposed by them.

In Bernstein's paper, he has proposed several countermeasures. He is mostly concerned about constant time AES implementations. In order to achieve this task he has proposed many hardware level and software level solutions and discussed many practical problems of them. Finally, he has stated that it cannot be guaranteed that the fetched T-table values will exist in the cache until the encryption process ends. Therefore, it is very difficult to achieve constant time AES. Paul Kocher stated that this constant time approach is good for masking timing characteristics. He targeted the discussion on Diffie-Hellman, RSA and DSS systems. He also stated that techniques used for blinding signatures could also be adapted to prevent attackers from knowing the input to the modular exponentiation function. David Brumley and Dan Bosh stated that most widely accepted defence against timing attacks is to perform RSA blinding. Secondly, they stated that making all RSA decryptions not dependent upon the input ciphertext. Besides this, they have proposed that making all RSA computations to be quantized would be another alternative. In Tromer et al. several countermeasures are proposed. They have stated that since these methods have different trade-offs and architecture and application dependency, a single recipe for all implementers cannot be recommended. Avoiding memory access, data independent memory access pattern, cache state normalization and process blocking and disabling cache access are some of advanced countermeasures they have proposed. To achieve them mostly the kernel support is needed. In this paper, we have implemented and tested some basic ideas extracted from the above researches. We have implemented them in a real system and tested their performance.

### III.   REVISITING BERNSTEIN'S ATTACK

Our attack environment is setup as demonstrated by Bernstein [4]. What Bernstein proposed is a technique to reduce the AES keyspace and then to perform brute-force to find the final key. The key-space reduction was made possible through the timing information collection.

We collected timing measurements on how long it takes to perform brute-force search on various sizes of key combinations so that we could use it to see the feasibility of the attack when the reduced key combinations are obtained from the timing information as in Bernstein attack. We have used Intel Pentium Dual Core, 2.1GHz processor for the timing measurement and the results are reported in Table I.

TABLE I
TIME TAKEN TO PERFORM BRUTE-FORCE SEARCH ON THE REDUCED KEY SPACE

| Key Space | Time for brute-force search (s) |
|---|---|
| $10^2$ | 0.01 |
| $10^3$ | 0.02 |
| $10^4$ | 0.02 |
| $10^5$ | 0.02 |
| $10^6$ | 0.09 |
| $10^7$ | 0.53 |
| $10^8$ | 4.58 |
| $10^9$ | 40.23 |
| $10^{10}$ | 348.23 |
| $10^{11}$ | 2977.83 |
| $10^{12}$ | 24512.69 |

We used the data in Table I to plot the graph shown in Figure 2. It was a linear graph with a gradient of $2.4 \times 10^{-8}$ using which we obtained Equation (1) to estimate the time required in second for brute-force searching in the particular machine.

$$SearchTime = 2.4 \times 10^{-8} \times keyspace \quad \text{---} \quad (1)$$

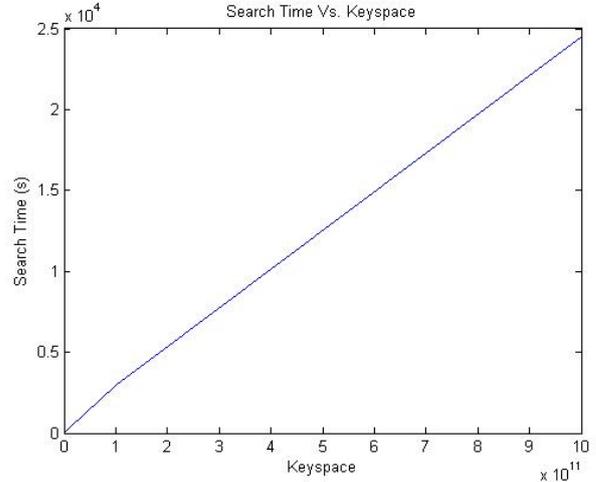

Fig. 2. Search Time verses the Key-space

### IV.   COUNTERMEASURES

In order to prevent cache timing attacks there are two major approaches to be followed. One is masking the leaking timing information and the other one is stopping information leakage. Second approach is very difficult to be implemented. Therefore, we took the first approach. In order to mask the leaking timing information, some disturbances should take place within the useful work. Therefore, there are several code fragments added into the AES implementation without changing its semantics. We tried to do these disturbances without heavily reducing the original efficiency of AES and they are described below.

#### 1. Random 'for' loops

As a countermeasure against cache-timing attacks, a random for loop was included in the AES implementation. First, a random number between 0 and 20 was generated and a 'for loop' was run from zero to the generated number. Algorithm 1 shows the code fragment implemented.

The *random_loop()* function in Algorithm 1 was called in *AES_encrypt* function. Before every encryption, random number is generated and random loop is run. This changes the encryption time significantly. Generating a random number spends nearly 3800 clock cycles and running a random for loop up to 20 iterations may spend about 100 or 200 clock cycles in the 733MHz server machine.

After using the changed AES implementation in the server and performing the attack, it is observed that some of the key bytes were missing. Because of incorrect timing information received from the server, the attacker could not correctly identify the actual timing pattern.

```
/** Generate random number between 0 and 20*/
int gen_random(){
        int iseed = (int)time(NULL);
        srandom (iseed);
        return rand()%20;
        }
/** Random for loop**/
int random_loop(){
        int x=gen_random();
        int i=0;
        int cnt=0;
        for(i=0;i<x;i++){
                cnt++;
                }
        return 0;
        }
```

Algorithm1. Random 'for' loops

When measuring the average number of cycles for encryption of 800-byte packets, it takes around 5062 clock cycles for original AES implementation while this changed AES consumes around 9303 clock cycles. This is 1.84 times higher than the number of cycles taken for unprotected AES implementation. It is observed that generating a random number consumed a long time compared to running a loop.

*2. Specified 'for' loops*

Random 'for' loops is a successful countermeasure. However, generating a random number consumes large number of CPU clock cycles and hence the AES encryption becomes slower. Therefore, the concentration is on a method based on loops but not generating a random number.

Here number 1777 is used as an initial number (there is no specific reason to choose this number). Then this number is divided by 17. The rounded answer is 104 and a 'for' loop is run from 0 to 104 at the beginning of an encryption. Then number 104 is divided by 17 and the rounded result is 6. Then 'for' loop is run from 0 to 6 times at the beginning of next encryption. Then number 6 is divided by 17 and the rounded result is 0. When the result is less than 6 the initial number 1777 is taken again. Likewise, the above steps were repeated.

A 'for loop' of 104 and then a 'for loop' of 6 is run repeatedly until encryptions occurring. Besides this, the division operation is also affecting the runtime of the encryptions. Algorithm 2 is the code fragment of the implementation.

The above *for_loop()* function was called in *AES_encrypt* function. Using this changed AES version some of the key bytes could be hidden. When measuring the average number of cycles for encryption of 800-byte packets, it takes around 5062 clock cycles for original AES implementation while this changed AES consumes around 5599 clock cycles. This is 1.11 times higher than the number of cycle needed for unprotected AES implementation.

```
static int gen = 1777; // initial number
int for_loop(){
        gen =(int)gen/17;
        int i=0;
        int cnt=0;
        if(gen<6)
                gen=1777;// initial num
        else{
                for(i=0;i<gen;i++){
                        cnt++;
                        }
                }
        return 0;
        }
```

Algorithm 2. Specified 'for' loops

*3. Pre-fetching T-table values*

This is another way of masking the leaking timing information. Here, we used four, 16-element arrays to pre-fetch the values from four T-tables. Before the encryption, some values of T-tables are read into those arrays. Therefore, the T-table values are loaded into the cache memory before the encryption.

However, we cannot guarantee that those values remain in the cache until the encryption process finishes. Because they can be removed from the cache when other needed data are loaded into the cache. However, when fetching some T-table values may be loaded into the cache before the encryption. It may cause to change the pattern of cache hits and misses when compared to the unprotected AES implementation. Because of this pre-fetching some needed T-table value can be loaded into the cache which value may not be loaded into the cache when using the unprotected AES. Otherwise, because of this pre-fetching some value may be evicted from the cache to provide room for newly loading values.

```
/*******four arrays to pre fetch********/
        u32 pre_fetch1[16];
        u32 pre_fetch2[16];
        u32 pre_fetch3[16];
        u32 pre_fetch4[16];
/** integers to manage array indices***/
        static int end_ind=16;
        static int start_ind=0;
```

Algorithm 3. Pre-fetching tables

This method is quite different compared to the two previous methods. Methods 1 and 2 did not change the pattern of cache hits and misses. Instead, they simply add some randomness to the encryption process. However here it changes the original pattern of cache hits and misses while adding some extra time consuming process into the encryption process. Some extra time consumption is to read the T-table values into an array. In the implementation we have used a 'for loop' for this and it will add some little extra time consumption in the encryption.

Instead of 16 element arrays, we can use larger sized arrays. However, it may reduce the efficiency and most of pre-fetched T-table values can be removed from the cache since the cache has limited space. We can do some experiment with the optimal size of those arrays. Algorithm 3 is the code fragment of the implementation.

AES_encrypt function is included in Algorithm 4.

---

```
…..
int i=0;//pre fetching for loop
r = key->rounds >> 1;
for (;;) {
/**********pre fetching*********/
        for(i=start_ind;i<end_ind;i++){
                pre_fetch1[i]=Te0[i];
                pre_fetch2[i]=Te1[i];
                pre_fetch3[i]=Te2[i];
                pre_fetch4[i]=Te3[i];
                }
        asm("nop");
/******************************/
        start_ind+=16;
        end_ind+=16;
        if(end_ind>256)
                end_ind=16;
        if(start_ind>192)
                start_ind=0;
……..
```

Algorithm 4. Pre-fetching

When measuring the average number of cycles for encryption of 800-byte packets, it took around 5062 clock cycles for unprotected AES implementation while the changed AES consumes around 5649 clock cycles. This is 1.12 times higher than the number of cycles in the original AES implementation.

### 4. Cache Partitioning

Another good way of preventing attack is cache partitioning. Here, we have allocated cache locations to load T-table values. Because of this allocated locations particular set of T-table values are loaded into a particular location. Therefore, those locations are not overwritten by other data, which are loaded into the cache. This may change the original pattern of cache accessing when an encryption is performing. In order to partition the cache the programmer must have understanding of the cache size of that specific machine. Algorithm.5 shows the code fragment of cache partitioning.

When measuring the average number of cycles for encryption of 800-byte packets, it takes around 5062 clock cycles for original AES implementation while this changed AES consumes around 3015 clock cycles. This is 0.60 times higher than the number of cycles in the unprotected AES implementation. This is very different compared to the other countermeasures. Time taken to encrypt the altered AES is lower than the unprotected AES. This might be because there are allocated memory areas to store T- table values and the values might be residing in the cache for longer time and it might have reduced the encryption time. Since most of the T-table values are available in the cache for the entire process cache miss rate may be reduced. Therefore, that might change the original pattern of encryption in unprotected AES and hence values may be missed.

---

```
….
static const u32 Te0[256]__attribute__((aligned(0x10)))=
{…...};
static const u32 Te1[256]__attribute__((aligned(0x1000))) =
{…..};
static const u32 Te2[256]__attribute__((aligned(0x10000))) =
{…..};
static const u32 Te3[256]__attribute__((aligned(0x100000)))
= {…..};
static const u32 Te4[256]__attribute__((aligned(0x1000000)
)) = {…..};
……...
```

Algorithm 5. Cache partitioning

Performing more experiments on this countermeasure to verify its behaviour is proposed as a future work.

### V. PERFORMANCE COMPARISON

If a countermeasure can miss a single key byte and the attacker knows the position where the byte is missed, the attacker has to perform a search on 256 times larger key space since there are 256 possible values for each byte position. Therefore, it makes it difficult to attack even the attacker knows the position where the byte is missed. However, if that clue is also unavailable to the attacker, it makes the attack even harder. If few key bytes were missed the key space would be larger in some power (number of bytes missing) of 256. Therefore, missing a key byte makes the key space larger and it makes the searching harder.

As we have reported in the previous section, different countermeasures have shown different amount of overheads. These overheads directly affect the efficiency of the AES implementation. If we look at the number of byte values missed for a particular countermeasure, it will also help us to compare them with each other. Therefore, we took average number of missing bytes in each countermeasure. Table II shows the performance information of each countermeasure. These are measured using 800-byte packets. Column two of the table lists the number of key bytes missing in the reduced keyspace after the attack. This is out of the total key 16bytes. The more the missing values, the better the countermeasure is. Column 3 is the number of clock cycle required for the countermeasure and the last column is the overhead of the countermeasure as a number of times compared to the original.

TABLE II
PERFORMANCE INFORMATION OF EACH COUNTERMEASURE

| Countermeasure | Avg. no of missing key bytes (m) | Avg. no of clock cycles per encryption (c) | X times slower than original AES (s) |
|---|---|---|---|
| Random 'for' loops | 7 | 9303 | 1.84 |
| Specified 'for' loops | 8 | 5599 | 1.11 |
| Pre-fetching | 10 | 5649 | 1.12 |
| Cache partitioning | 14 | 3015 | 0.60 |

Figure 5 and 6 show the graphical representation and therefore a comparison of the results.

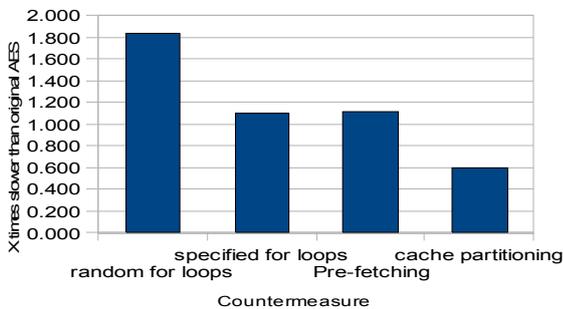

Fig. 5. Graph of countermeasures and information about how much they are slower than the original AES

Using this information, we propose Equation (2) to measure efficiency of the countermeasures tested. It would give an estimate about the efficiency of the countermeasure. Where *m* is the number of missing values and *s* is the performance.

$$Efficiency = \frac{1}{s} m \quad \text{--------- (2)}$$

When using this equation we have to consider boundary cases. That is, when the efficiency is zero that implies either *m* is zero or *1/s* is zero. *1/s* is zero means *s* is very large and hence the implementation is very slow. If *m* is zero, it means all the key bytes are available in the potential key space and still there is a possibility to search the correct key (may be vulnerable to attack). However, if most of the values from 0-256 are available in all 16 key byte positions, performing a search is impossible (strong defence). Therefore, when efficiency is zero we have to consider the boundary cases discussed here.

According to the equation derived, we have measured efficiencies of above four countermeasures.

Efficiency (Random 'for' loops)     = 3.80
Efficiency (Specified 'for' loops)  = 7.20
Efficiency (pre-fetching)           = 8.93
Efficiency (Cache partitioning)     = 23.33

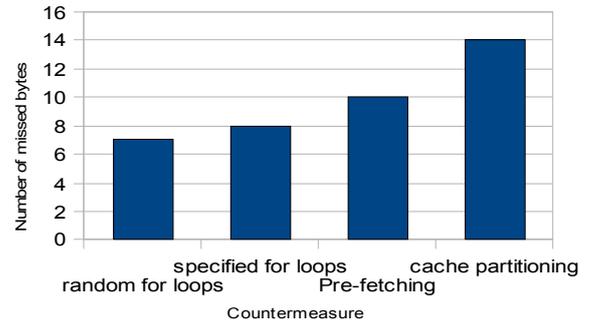

Fig. 6. Graph of countermeasures and number of missing key bytes

## VI. CONCLUSION

In this paper, we have designed and implemented four countermeasures against the remote cache timing attack proposed by Daniel Bernstein. The countermeasures are compared for their efficiency and overheads. Besides, we have briefly considered a way to quantify their efficiency and have proposed an estimate using available parameters.

REFERENCES

[1] Wikipedia, (2011,May). *Advanced Encryption Standard* [Online]. Available: http://en.wikipedia.org/wiki/Advanced_Encryption_Standard
[2] Wikipedia, (2011,May). *Side Channel Attacks* [Online]. Available: http://en.wikipedia.org/wiki/Side_channel_attack,
[3] Andrew S. Tanenbaum, "Computer Hardware review," in Modern Operating Systems, 3rd Ed. New Jersey: Prentice Hall, 2008.
[4] Daniel J. Bernstein, "Cache Timing Attacks on AES", April 2005.
[5] Paul Kocher. "Timing attacks on implementations of Diffie-Hellman, RSA, DSS and other systems", Proceedings of the 16th Annual International Cryptology Conference on Advances in Cryptology (CRYPTO 1996), p 104-113, Springer-Verlag London, UK, 1996.
[6] D.Brmley and D.Bosh. "Remote timing attacks are practical", in USENIX, August 2003.
[7] Edward W. Felten and Michael A. Schneider. "Timing Attacks on Web Privacy", in Proceedings of the 7th ACM conference on Computer and communications security, New York, NY, USA, 2000.
[8] Eran Troman, Dag Arne Osvik and Adi Shamir, "Efficient Cache Attacks on AES, and Countermeasures", *Journal of Cryptiology*, J.Cryptol.(2010) 23:37-71 DOI:10.1007/s00145-009-9049-y.
[9] Darshana Jayasinghe, Jayani Fernando, Ranil Herath, and Roshan Ragel, "Remote Cache Timing Attack on Advanced Encryption Standard and Countermeasures", in Proceedings of the 5th International Conference on Information and Automation for Sustainability (ICIAfS2010), pp 177-282, Colombo, Sri Lanka, December 2010.